# A Coordination-Based Model for the Prediction of Surface Energies and the Shape of Metal Particles

Shyama Charan Mandal[‡,†] and Frank Abild-Pedersen[†,*]

[‡]SUNCAT Center for Interface Science and Catalysis, Department of Chemical Engineering, Stanford University, Stanford, California 94305, United States

[†]SUNCAT Center for Interface Science and Catalysis, SLAC National Accelerator Laboratory, Menlo Park, California 94025, United States

*Email: abild@slac.stanford.edu

**Abstract**

Surface energies of metal-based systems are important for determining the Wulff-constructed shapes of metal nanoparticles and understanding the stability. We have developed a coordination number-based model to predict the total energy of metal-based systems across a wide range of configurations. Our model has been tested against Density Functional Theory (DFT) calculations for late transition metals. This method enables on-the-fly surface energy predictions and allows for the Wulff construction of metal particles for a random number of elemental atoms and without the need for DFT calculations. By making a division between atoms in the different layers of the model system we can considerably improve the accuracy of the model, suggesting a dissimilarity between the electronic structure due to an alternating compression and expansion of atomic layers in the near-surface region. We find that our model accurately and effectively provides valuable insights into the distribution and stability of nanoparticle surfaces.

**Keywords**: density function theory; coordination-based ε-scheme model; surface energies; Wulff construction; nanoparticle

**Introduction**

Metal nanoparticles are highly valuable and have extensive applications in various fields such as drug discovery, biological detection, chemical catalysis, and more.[1–3] The usefulness arises from their unique properties which differ significantly from those of bulk materials and atomic-sized counterparts. The properties of metal nanoparticles are influenced not only by their orientations but also by their size and shape.[4,5] Therefore, a deep fundamental understanding is required to engineer these properties for desired applications. The initial understanding of metal nanoparticles began with the hypothesis by Wulff, that the surface energy along a given crystal orientation is proportional to the length of the vector and therefore effectively provides a framework for the construction of the equilibrium shape of particles.[6–8] A rigorous mathematical explanation for this conjecture was given by von Laue[9] and later proven in a more sophisticated manner by Dinghas.[10] Although extensions, such as the Winterbottom[11] and Summertop models,[12,13] have been proposed, the Wulff construction method is still widely used and has remained largely unchanged over the past few decades.

Supported metal nanoparticles are generally used in heterogeneous catalysis for a variety of reactions[14–17] and the equilibrium shape of the catalyst in an inert environment can be determined using the Wulff construction method. To understand how the particle shape of a metal catalyst evolves in a chemical environment, one must first identify the structure and shape as defined by the electronic arrangement of the elemental metal. The vast number of atomic configurations in metal nanoparticles[18] makes this task very challenging. Hence, it is imperative to develop a model based on physical principles that enables an accurate prediction of input quantities to the Wulff construction method. The Wulff construction depends on surface energies as input and constructs an arrangement of N atoms that minimizes the total surface energy within a fixed volume set by

N. The surface energy (γ) is a scalar function and the resulting Wulff shape is composed of orientations with low surface energy. The resulting particle exhibits a faceted morphology composed of planes, edges, and corners, which reflect the crystalline nature of metals. Even though obtaining the Wulff shape is a simple process, it still requires quantifying the energy for a large selection of surfaces. These energies can be obtained using DFT, but the work is tedious, and the computational cost increases significantly for higher index surfaces. Recently, we have developed a model to describe the metal binding energies of active surface sites. The model is based on a set of statistically averaged energy parameters that carry information of the local environment of the active site.[19] The method is very accurate for quantifying surface bond strengths and estimating energy differences for single atom transitions, however, it fails to recover cohesive energies and hence it is not well suited to describe total energies and surface energies. In this article, we introduce a simple model that better capture the sudden changes in the coordination environment allowing for an accurate and efficient prediction of total energies and surface energies thus enabling for the Wulff construction of nano particles of metal-based systems. We verify the model accuracy for various metals; Ag, Au, Co, Cu, Ir, Ni, Os, Pd, Pt, Rh, and Ru in the face-centered cubic (fcc) crystal structure and demonstrate an accurate and efficient model based on coordination number to understand the behavior of different metals.

**Results and discussion**

**Prediction of total energies**

The surface energy (γ) can be expressed as $\gamma = \frac{1}{2A}(E_{slab} - NE_{bulk})$. In the equation, A is the area of the one exposed surface, $E_{slab}$ is the total energy of the considered slab, N is the number of metal atoms present in the slab and $E_{bulk}$ is the total energy of one bulk metal atom. As per equation, obtaining the total energy of a surface slab is crucial for quantifying its surface energy. Therefore, our primary focus is on predicting the total energies of various surface orientations correctly. Our recently developed coordination-based α-scheme model[19] has been effective in accurately predicting metal binding energies, which are valuable for various catalysis-related applications, such as predicting adsorbate binding energies,[20] classifying hydrocarbon-based adsorbates,[21] and exploring correlations among different metals and alloys[22]. The α-scheme model is based on bond strengths within different coordination environments and is therefore very accurate for reactions with a high degree of anisotropy (Figure 1a) like when quantifying single atom bond energies and estimating energy differences for single atom transitions, however, it fails to recover reaction with lower degree of anisotropy like cohesive energies. From the bond energy picture, the cohesive energy of an atom is represented by the addition of all the α's for a given coordination number, cn, ($E_{coh} = \sum_{i=1}^{cn} \alpha_i^Z$). Because of the isotropy related with the cohesive energy all of the $\alpha_i^Z$ values will be the same, $\alpha_1^Z = \alpha_2^Z = \cdots = \alpha_{12}^Z$, for an atom in the bulk and the degree of isotropy decreases when the atom coordination decreases. We do not expect the α-scheme model to be accurate when quantifying cohesive energies and surface energies since it explicitly includes geometric and electronic effects induced by the changes in the coordination environment and relaxation in the d-electronic states due to structural and electronic relaxation and therefore $\sum_{i=1}^{cn} \alpha_i^Z$ does not agree with the $E_{coh}$ as the $\alpha_i^Z$ values are different for different coordination environments (Figure 1a). Here, we have developed a new, simple model (ε-scheme model) based on the coordination number of each metal atom, using a similar approach. We quantify the total cohesive energy, $E(n, Z)$, of any metal system with atomic number Z having n number of atoms with varying coordination as $E(n, Z) = \sum_{i=1}^{n} \varepsilon_i^Z$, where $\varepsilon_i^Z$ represents the cohesive energy of each metal atom in its corresponding coordination environment. To parametrize the coordination number-based energies, $\varepsilon_i^Z$, for all the metals considered, we performed a series of DFT calculations. We examined various surfaces (100, 110, 111, 210, 211, 221, 310, 311, and 320) with different coordination environments, as well as a bulk 3×3×3 structure to obtain the parameterized $\varepsilon_i^Z$ values for each metal atom in different coordination environments. In total, 10 structures were optimized and used to parameterize the ε-scheme model. To illustrate how the model is parameterized, let's consider an N×N surface slab of L layers in the (hkl) direction of an fcc crystal structure with n atoms of atomic number Z. If we let $n_s$ represent the number of metal atoms in the surface layer of the model slab, $cn_i$ be the coordination of the $i^{th}$ atom, and we treat all the atoms in the remaining L-2 layers of the slab as bulk atoms then the total energy of the system, $E[n]$, can be represented as $E[n, Z] = 2\sum_{i=1}^{n_s} \varepsilon_{cn_i}^Z + \sum_{j=1}^{L-2}\sum_{k=1}^{n_s} \varepsilon_{12_{bulk}}^Z$ (Figure 1b). It has been shown in an earlier study that even though the atoms below the surface layer has coordination number 12 they behave very differently from their bulk counterparts due to second order effects induced by the lower coordination of the atoms in the surface layer.[19] To further increase the accuracy one can expand the set of $\varepsilon_i^Z$ values with an additional parameter $\varepsilon_{12_{ss}^{hkl}}^Z$ (ss represents subsurface) to account for the effect along the hkl surface direction such that $E[n, Z] = 2\sum_{i=1}^{n_s} \varepsilon_{cn_i}^Z + 2\sum_{j=1}^{n_s} \varepsilon_{12_{ss}}^Z + \sum_{k=1}^{L-4}\sum_{l=1}^{n_s} \varepsilon_{12_{bulk}}^Z$ (Figure 1c).

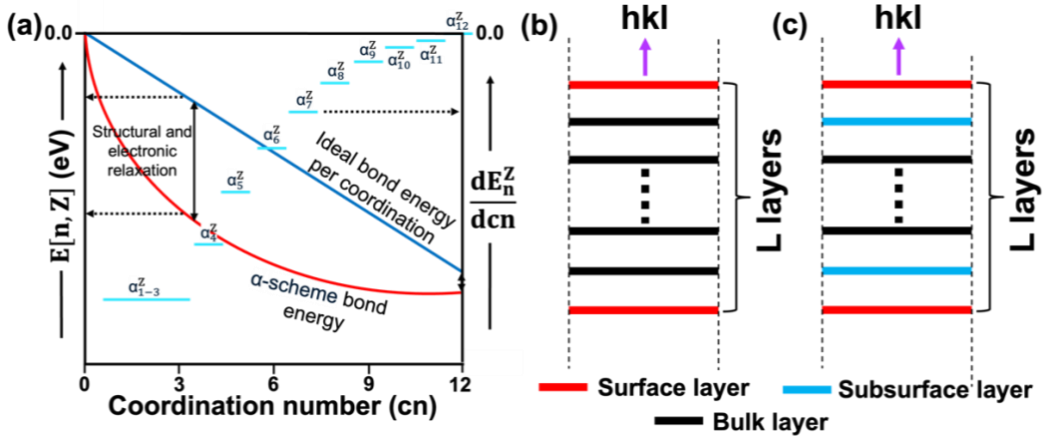

**Figure 1**: (a) Schematic representation of ideal bond energies per coordination and bond strength based on the α-scheme model. The differences in the α-scheme model's bond energies are due to the structural and electronic relaxation. The ideal bond energies per coordination leads to the cohesive energy whereas in the case of α-scheme model's bond energies leads to the corresponding α ($\alpha_i^Z$ is the α value at i$^{th}$ coordination environment of metal Z). (b) Side view of the model fcc(hkl) surface with L layers without considering subsurface and bulk separately, (c) Side view of the model fcc(hkl) surface with L layers with considering subsurface and bulk separately. Here, cn represents the coordination number whereas s, ss and bulk represents surface, subsurface and bulk metal atoms, respectively. $E(n, Z)$ denotes the total energy of the system with atomic number Z, while $\varepsilon_i^Z$ represents the energy of metal atom in the i$^{th}$ coordination environment.

This approach has been applied to all surfaces considered in our study (Figure 2). The considered surfaces and bulk structures with their corresponding coordination number and statistics, are listed in Table S1. We note that surfaces relevant to catalysis does not expose atoms with coordination lower than 6 and they were therefore excluded from this study. Hence, the energies associated with coordination numbers 1 through 5 are collectively included in the energy for coordination number 6, denoted as $\varepsilon_{1-6}^Z$, such that $\varepsilon_{1-6}^Z = \varepsilon_1^Z + \varepsilon_2^Z + \varepsilon_3^Z + \varepsilon_4^Z + \varepsilon_5^Z + \varepsilon_6^Z$.

Depending on whether we exclude the subsurface corrections or not, there are seven or eight parameters $\varepsilon_i^Z$ to optimize. Hence, we need at least eight equations with sufficient statistics for each parameter to determine an accurate representation of the $\varepsilon_i^Z$ values. We have considered 9 surfaces together with the bulk structure (Figure 2) a total of 10 equations which is sufficient to optimize the set of $\varepsilon_i^Z$'s. The inclusion of more structures will lead to more accurate $\varepsilon_i^Z$ values for the broader range of surfaces. The $\varepsilon_i^Z$ value parameterization using the 10 equations was carried out in two ways: in one case, no constraints were applied, while in the other, we imposed the constraint that the energy of $\varepsilon_i^Z$ should increase with increasing coordination numbers. We impose the constraint based on the d-band structure as a function of coordination. Therefore, there are four possible $\varepsilon_i^Z$ optimizations for every metal such as (i) unconstrained optimization without subsurface corrections, (ii) constrained optimization without subsurface corrections, (iii) unconstrained optimization with subsurface corrections, and (iv) constrained optimization with subsurface corrections.

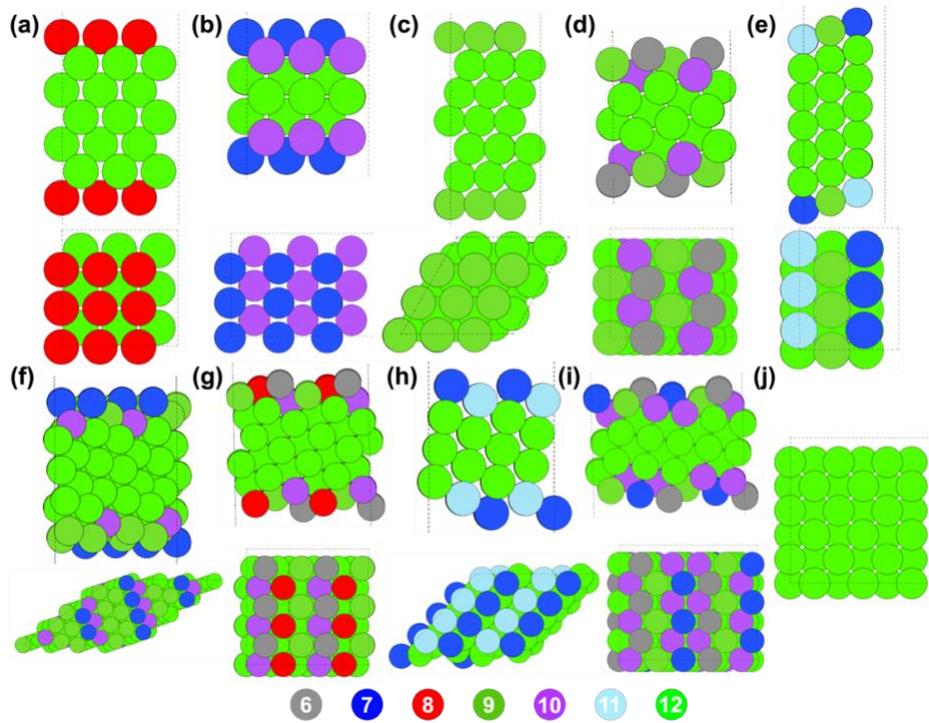

**Figure 2**: Considered structures for the parameterization of coordination-based $\varepsilon_i^Z$ in the ε-scheme model. (a) 100, (b) 110, (c) 111, (d) 210, (e) 211, (f) 221, (g) 310, (h) 311, (i) 320 surfaces and (j) bulk fcc 3×3×3 structures. The corresponding coordination numbers with their colors have been shown.

The optimized values of all the constrained $\varepsilon_i^Z$ for each metal considered have been shown in Figure 3. The optimized $\varepsilon_i^Z$ values without constraints have been plotted in the supporting information Figure S1 and all numeric values are listed in Table S2 and S3 of the supporting information. The observed unconstrained $\varepsilon_i^Z$ values do not follow a discernible trend. This arises from the presence of only one type of atom in the bulk structure, resulting in a fixed $\varepsilon_{12}^Z$ value. Consequently, the parameterization of other $\varepsilon_i^Z$ values is influenced by this fixed bulk $\varepsilon_{12}^Z$ values. However, we have considered constrained model based on the d-band structure as a function of coordination as the more neighbors of a metal atom leads to lower its d-band energy.[23]

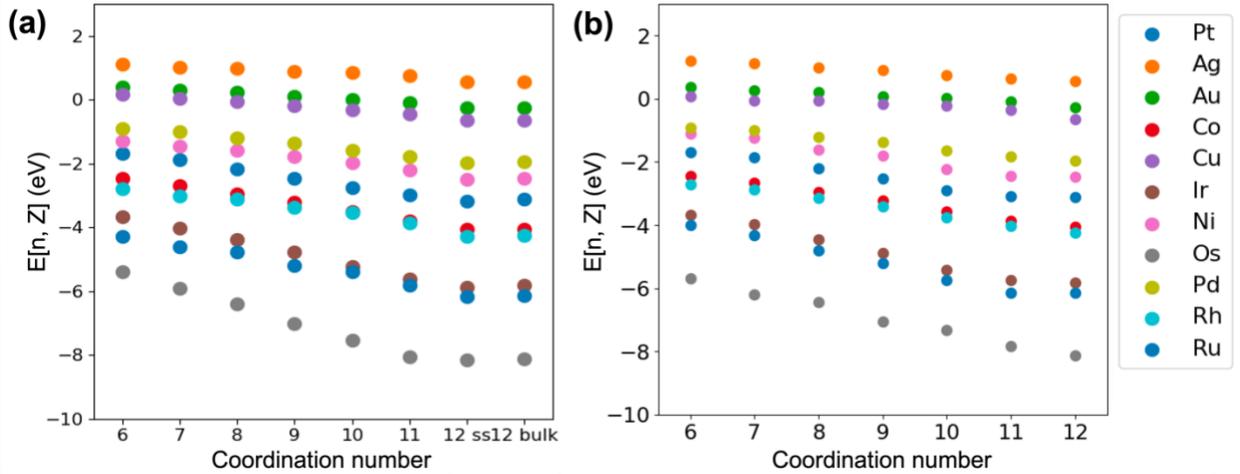

**Figure 3**: The considered fcc metal atoms (a) constrain $\varepsilon_i^Z$ energies with considering subsurface and bulk metal atoms separately, and (b) constrain $\varepsilon_i^Z$ energies without considering subsurface and bulk metal atoms separately where M represents the considered metal and ss represents the subsurface metal atoms having coordination number 12.

Once all $\varepsilon_i^Z$ values were optimized under the various conditions we compared the total DFT calculated energies with the ε-scheme model predicted total energies shown in Figure S2-S5 in the supporting information. The mean absolute errors (MAE) obtained have been listed in Table S4.

Across the board the predicted total energies whether using constrained or unconstrained $\varepsilon_i^Z$ shows negligible MAEs compared to the DFT calculated values. There is an indication that predictions for the metals Ru, Pd, Ir, Os, and Pt show elevated MAEs but the errors are insignificant for our purposes. In general, the obtained results show that the ε-scheme model is very accurate and efficient for the prediction of total energies of various surfaces and bulk structures based on the coordination environment and the identity of metals.

**Prediction of surface energies and Wulff shapes**

With an accurate model prediction of the total energy of the surfaces and bulk structures, we can move forward with the quantification of the surface energies. As shown earlier, the energy of a surface along a specific crystallographic orientation defined by the Miller indices (hkl) is a function of the total energy of the surface, the bulk energy, and the surface area. When evaluating the surface energies using the model predictions, we again consider the effects introduced by the subsurface corrections and the physical constraints set by the d-band behavior at varying coordination. In Figure 4 we have shown the comparison between the surface energies calculated using DFT and the predicted surface energies using the d-band constrained ε-scheme with and without subsurface corrections. The unconstrained comparisons are shown in Figure S6 of the supporting information. All the MAEs have also been listed in Table S5 of the supporting information.

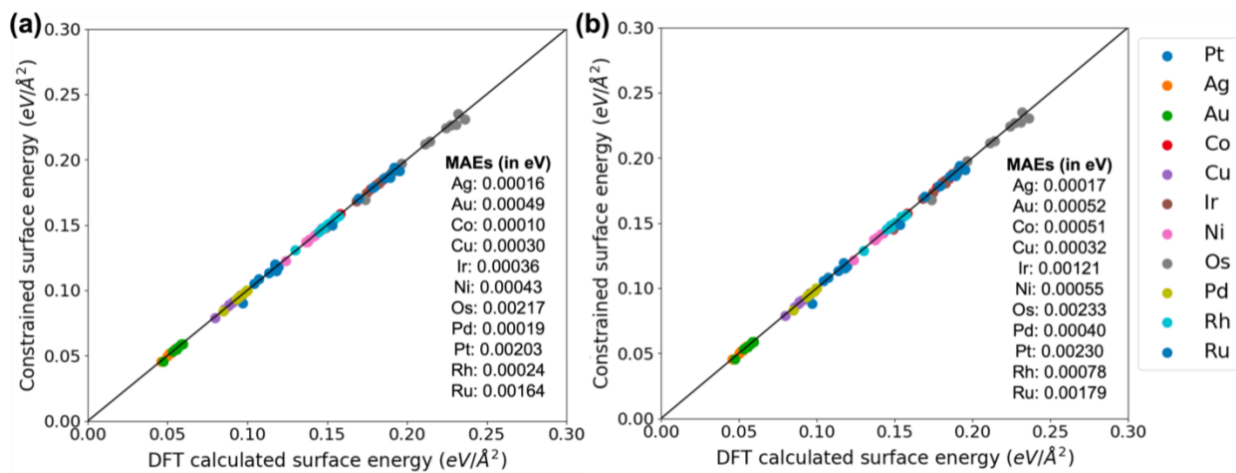

**Figure 4**: Predicted surface energies using the constrained parametrization of the ε-scheme physical with (a) inclusion of subsurface corrected $\varepsilon_{12}^Z$ values and (b) without subsurface corrections to the $\varepsilon_{12}^Z$ values vs. DFT calculated surface energies. MAEs are shown in the figures.

Overall, the model is very accurate in predicting the surface energies of all metals considered. MAEs are only weakly affected by applying the constraints whereas introducing additional parameters to account for the enhanced compression and associated electronic effects in the surface layer leads to minor improvements in the accuracy. In Table S6 of the supporting information we have compared our model data from Figure 4a with similar data reported by others[24] and we find our data to be in good agreement with these earlier reports. Based on this, we are confident that our model predictions can be used to explain surface structural properties of catalyst nanoparticles as well as extended surfaces. The simplicity of the ε-scheme enables us to evaluate surface energies on the fly and hence obtain the Wulff shape, the lowest energy structure of nanoparticles, for any ensemble N atoms. In the following, the predicted surface energies have been used with the software package WulffPack[25] to obtain the Wulff shapes of an N atom ensemble in the fcc crystal structure for the metals considered in this study. In addition, due to the flexibility and simplicity of the model, we have expanded the surface space considered to include some of the higher index surfaces, such as 321, 322 and 332 to obtain more accurate representations of the particle shapes. The particle shapes for the individual metals are shown in Figure 5.

A closer examination of the particle shapes allows us to address various catalytic relevant properties such as the distribution of crystallographic planes, particle volume, surface area, fraction of catalytically relevant surface sites, the total surface energy, the average surface energy, the strain energy, and the relative frequency of corners and edges on the nanoparticle. The metals that display more heterogeneity in the Wulff shapes are those where surface energies fall within a narrower energy window.

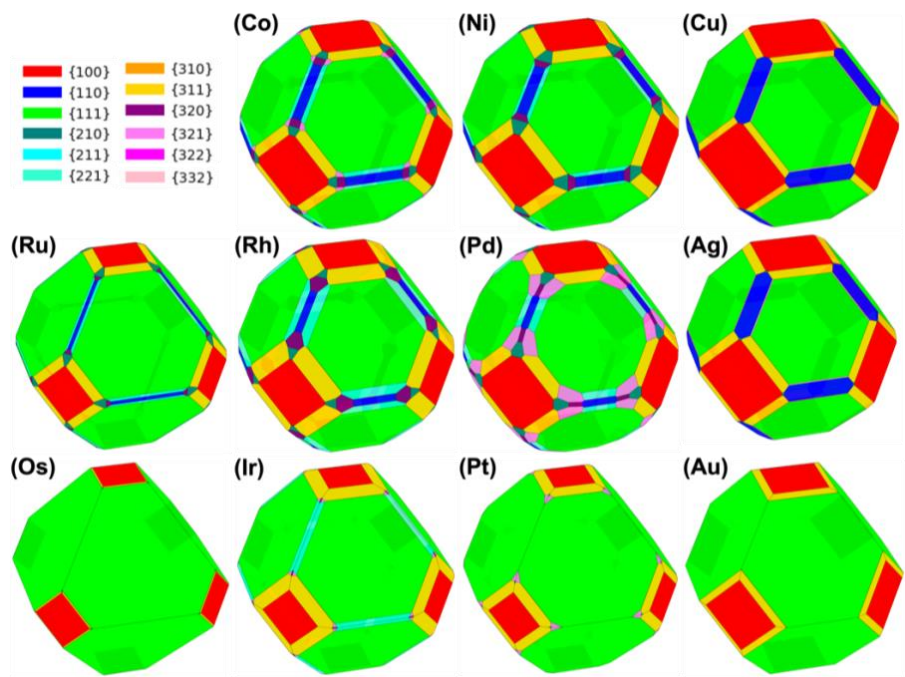

**Figure 5**: Predicted lowest energy particle shapes of an ensemble of N atoms for all the metals considered based on surface energies using the ε-scheme.

**Conclusion**

We have developed a model (ε-scheme) based on a set of simple parameters that carry information of metal-metal coordination and energy that efficiently and accurately allows for the prediction of surface energies. At first, we have focused on the prediction of total energies based on the coordination number of every metal present in the considered structures. To build our model, we have considered 3d (Co, Ni and Cu), 4d (Ru, Rh, Pd and Ag) as well as 5d (Os, Ir, Pt and Au) metals in the fcc crystal structure. The parametrization based on 10 DFT calculations on various surfaces and bulk structures and the model parameters are then used to predict total energies and surface energies. We find that by introducing a slight correction to atoms in the subsurface we can improve on the accuracy of the model suggesting a physical dissimilarity between subsurface and bulk metal atoms. To further validate our model, we have used our model surface energies to predict the Wulff shapes of an ensemble of N atoms of all the considered metals which compares well with shapes reported in the literature. Hence, our model provides an accurate and efficient way to determine particle shapes without any further DFT calculations which is important when evaluating catalytically relevant properties.

**Methods**

**Computational Details**

All the density functional theory (DFT) calculations have been performed using Vienna Ab initio Simulation Package (VASP) with the Atomic Simulation Environment (ASE).[26–29] The BEEF-vdW functional has used to include exchange-correlation effects.[30,31] The BEEF-vdW functional has been proved for accurately describing metal surfaces earlier.[21,32] A cutoff energy of 500 eV was used during optimization. All the considered structures were optimized until the total energies converged to $10^{-5}$ eV, and the forces converged 0.05 eV/Å. The considered metal lattice constants were obtained from earlier study.[19,21,22,32] To diminish the periodic image interactions, more than 15 Å vacuum was introduced between the slabs in the z-direction. The reciprocal space of the surfaces, a Monkhorst-Pack method of 4 × 4 × 1 k-point grid was utilized within the Brillouin zone.[33] Furthermore, we have added dipole corrections to remove artificial periodic interactions between the slabs.[34] In the case of surface energy calculations, both the surfaces of the considered slab are alike. The surface energy (γ) can be expressed as follow.

$$\gamma = \frac{1}{2A}(E_{slab} - NE_{bulk}) \quad (1)$$

In the above equation 1, A is the area of the one exposed surface, $E_{slab}$ is the total energy of the considered slab, N is the number of metal atoms present in the slab and $E_{bulk}$ is the total energy of one bulk metal atom.

**Data availability**

The data used to analysis this study is available from the corresponding author upon request.

**Associated Content**

**Supporting Information:**

Statistics of coordination numbers for the considered surfaces, bulk structures, and fcc metal atoms with unconstrained $\varepsilon_i^Z$ energies. Unconstrained as well as constrained $\varepsilon_i^Z$ values for all considered metals with and without subsurface corrections. Unconstrained as well as constrained comparison of DFT calculated vs. ε-scheme model predicted total energies for all considered metals with and without subsurface correction. Mean absolute errors (MAEs) of the DFT calculated total energies and ε-scheme model predicted total energies, comparison of surface energies using unconstrained $\varepsilon_i^Z$ values vs. DFT calculated surface energies. MAEs of the DFT calculated vs. ε-scheme model predicted surface energies for all the considered metals, comparison of surface energies with earlier reports and ε-scheme model predicted values.

**Acknowledgement**

We acknowledge support from the U.S. Department of Energy, Office of Science, Office of Basic Energy Sciences, Chemical Sciences, Geosciences, and Biosciences Division, Catalysis Science Program to the SUNCAT Center for Interface Science and Catalysis. This research used resources of the National Energy Research Scientific Computing Center; a DOE Office of Science User Facility supported by the Office of Science of the U.S. Department of Energy under Contract No. DE-AC02-05CH11231 using NERSC award BES-ERCAP0024127. We acknowledge computational support from the SLAC Scientific Data Facility (SDF).

**Author contributions**

F.A.-P. conceived the idea. S.C.M. computed dataset and compiled the dataset. All authors have contributed to the writing of the paper and the analysis of the data.

**Competing Interests**

The authors declare no competing interests.

**Table of Contents Entry**

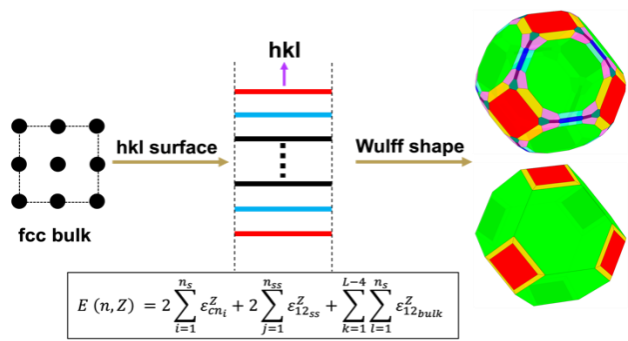